\documentclass[sigconf, nonacm]{acmart}

\usepackage{tikz}
\usepackage{amsmath}
\usepackage{subfigure}
\usepackage{subcaption}
\usepackage{enumitem}
\usepackage{url}
\usepackage{multirow}
\usepackage{comment}
\usepackage{xcolor}
\usepackage{graphicx}
\usepackage{pdfpages}
\usepackage{pifont}
\usepackage{fontawesome}
\usepackage{multicol}
\usepackage[normalem]{ulem}
\usepackage{booktabs}
\usepackage{makecell}
\usepackage{tabularx}
\usepackage{tabulary}

\usepackage{xspace}
\newcommand{\sysname}{DMG\xspace}

\AtBeginDocument{%
  }

\begin{document}

\title{\sysname: A Scalable and Efficient Memory-Disaggregated Graph Processing System}


\author{Yizou Chen}
\affiliation{%
 \institution{The Chinese University of Hong Kong}
 \country{}
}
\email{chenyz@cse.cuhk.edu.hk}

\author{Tsun-Yu Yang}
\affiliation{%
 \institution{Duke University}
 \country{}
}
\email{tsun-yu.yang@duke.edu}

\author{Zhisheng Hu}
\affiliation{%
 \institution{The Chinese University of Hong Kong}
 \country{}
}
\email{zshu23@cse.cuhk.edu.hk}

\author{Baotong Lu}
\affiliation{%
 \institution{Microsoft Research}
 \country{}
}
\email{baotonglu@microsoft.com}

\author{Ming-Chang Yang}
\affiliation{%
 \institution{The Chinese University of Hong Kong}
 \country{}
}
\email{mcyang@cse.cuhk.edu.hk}

\begin{abstract}
Traditional graph processing systems are built on monolithic servers, which couple a fixed ratio of compute and memory resources but often result in resource under-utilization in data centers.
Although the disaggregated memory (DM) architecture has emerged to address this inefficiency, we identify that existing graph processing systems on DM remain highly impractical.
They rely on unscalable architectures that fail to scale beyond a single memory node and a single compute node, 
and they require compute-side caches that are orders of magnitude larger than conventional practice in DM.

To this end, this paper presents \sysname, the first practical graph processing system on DM, which demonstrates superior system scalability and cache efficiency while delivering high performance.
To improve efficiency of graph retrieval on DM, \sysname proposes a \textit{DM-friendly graph store} with retrieval optimizations. 
To mitigate costly update propagation, \sysname presents an \textit{adaptive update coordinator} that coordinates compute and memory nodes to perform update propagation with low overhead.
To enable fast and effective load balancing, \sysname employs a \textit{two-stage workload manager} that includes a coarse-grained initial partitioning and a fine-grained runtime re-scheduling.
Experimental results substantiate that compared with the state-of-the-art DM-based graph processing system, \sysname can elastically scale up both compute and memory resources, delivering up to 4.9$\times$ better performance and accommodating graphs with ever-increasing sizes; meanwhile, it effectively tames the compute-side cache demands by up to 18.9$\times$, positioning itself as a DM-ready solution in practice.

\end{abstract}

\maketitle

\section{Introduction}

Graph is a powerful abstraction for modeling complex relationships among entities and is essential in various applications, from traditional data analytics tasks (social networks, bio-informatics, and e-commerce transaction networks) to modern AI-related tasks (vector search~\cite{HNSW,DiskANN,pipeann_osdi,hmann_nips}, recommendation systems~\cite{GNNRecomSurvey}, retrieval-augmented generation~\cite{graphrag}, and graph neural networks~\cite{MGG_OSDI,Dorylus_OSDI,knightking}).
Especially, graph processing systems~\cite{galois,ligra,gemini_2016,compressgraph_sigmod,10.1145/3709656} have been widely studied as they embody optimizations for fundamental access and computation patterns on graph-structured data.

Traditional distributed graph processing systems are built on monolithic servers~\cite{powergraph_2012,gemini_2016,gluon_2018}, where different resource components (e.g., CPU and DRAM) are tightly coupled, scaled at a fixed rate, and lack sharing across physical servers.
Such systems are inefficient in \textbf{\textit{resource utilization}} (detailed in §~\ref{sec:motivate_graph_on_DM}).
Specifically, to meet peak demand in either computation or memory capacity, these systems allocate multiple monolithic servers, leading to over-provisioning of the other resource.
Additionally, graphs need to be partitioned across isolated memories in traditional distributed systems, which require replicating boundary vertices. 

Recently, the disaggregated memory (DM) architecture has been proposed to improve resource utilization troubled by monolithic servers in data centers, attracting extensive attention from both academia~\cite{legoos_osdi,shard_vldb,Disaggregation_vldb25,dLSM,zhang2022optimizing} and industry~\cite{dm_moti_google,PolarDB-MP_sigmod24,GaussDB_vldb24,PolarCXLMem_sigmod}.
It decouples compute and memory components of monolithic servers into independent compute and memory pools bridged by fast and advancing interconnect technologies such as RDMA~\cite{rdma2023manual} and CXL~\cite{cxl2023manual}.
The compute pool consists of compute nodes (CNs) with many CPU cores but little DRAM as cache, while the memory pool consists of memory nodes (MNs)\footnote{We use server and node interchangeably: \textit{server} mainly refers to traditional monolithic server, while \textit{node} mainly refers to memory/compute node in DM.} with ample DRAM but few CPU cores.

Thanks to its distinctive architecture, DM offers great potential to achieve high resource utilization for graph processing.
Specifically, as compute and memory resources are decoupled, DM-based graph processing systems can on-demand provision compute and memory resources independently to meet diverse workload needs~\cite{FaaSGraph,famgraph}.
Additionally, the large, shared memory pool of DM enables the maintenance of a unified graph view, which allows CNs to access a single copy of vertex data, thereby avoiding replication overhead.

Unfortunately, we find that existing attempts for graph processing on DM~\cite{famgraph,fargraph} remain highly impractical, as they fail to meet the criteria of DM and fall far short of leveraging DM's benefits (§~\ref{sec:existing_work}).
First, they are impractical for the \textbf{\textit{unscalable system}} that ties execution to a single CN and a single MN.
They cannot accommodate large graphs that exceed one MN’s memory or use computation power more than one CN can provide, contradicting DM’s promise of elastic scaling.
Second, they are impractical for \textit{\textbf{violating DM convention} to over-demand compute-side cache}.
DM typically provisions limited cache per CN to retain independence in resource allocation, yet prior work~\cite{famgraph} reports cache requirements of more than an order of magnitude higher for the \textit{clueweb12} dataset on one CN. 
More critically, such demand grows linearly with graph size and replicates across CNs when scaling out, thereby re-coupling resource allocation, amplifying resource usage, and nullifying the resource efficiency enabled by disaggregation.

\begin{table}[t]
    \footnotesize
    \centering
    \setlength{\abovecaptionskip}{5pt}
    \label{tab:system_comparison}
    \caption{Comparison of different graph processing systems.}
    \renewcommand{\arraystretch}{1.2}
    \newcolumntype{C}{>{\centering\arraybackslash}X}
    \begin{tabularx}{0.98\columnwidth}{l | C | C | C}
        \toprule
         & \textbf{Traditional Distributed} & \textbf{Existing DM-based} & \textbf{\sysname (DM-based)} \\
        \midrule
        \textbf{Resource Utilization} & Bad & Good & Good     \\
        \textbf{System Scalability} & Good & Bad & Good \\
        \textbf{DM Convention} & -- & Bad & Good \\
        \bottomrule
    \end{tabularx}
    \label{tab:system_comparison}
\end{table}

As summarized in Table~\ref{tab:system_comparison}, motivated by the \textit{resource inefficiency} and \textit{impracticalities} of existing systems, we aim to enable high-performance graph processing on DM with superior system scalability and DM-conventional cache demand.
However, even with a sensible partitioning scheme (as discussed in §~\ref{sec:motivation_challenge}), it is non-trivial to realize high performance with the practical architecture that fully reaps the benefits of DM.
In particular, three key challenges remain:
\textbf{\textit{(1) Retrieving graphs from DM exhausts network.}} 
The randomness, dependency, and fine granularity of graph accesses generate numerous small remote requests, quickly saturating RDMA’s limited IOPS capacity and limiting performance.
\textbf{\textit{(2) Remote updates to DM are unavoidable yet costly.}}
The skewed connectivity in real-world graphs makes it impossible for partitioning to fully avoid edges targeting outside a CN’s cache~\cite{powerlyra_eurosys,NE_partition}. 
Although only a small fraction of edges trigger remote updates, they can still cause substantial slowdown since RDMA atomic operations are far more costly than local ones. 
\textbf{\textit{(3) DM urges fast and effective load balancing.}}
Existing graph partitioning methods for inter-CN balancing are time-consuming, making them impractical for DM where resources are frequently reconfigured. 
Within each CN, severe tail effects arise since hub-vertices with long edge lists amplify the impact of high-latency RDMA operations.

To address the above challenges, this paper presents \textbf{\sysname}, 
the first practical \textbf{\underline{G}}raph processing system on \textbf{\underline{D}}isaggregated \textbf{\underline{M}}emory, 
which demonstrates superior system scalability and cache efficiency while delivering high performance.
\sysname incorporates three key and innovative designs:
First, to achieve efficient graph retrieval, \sysname presents a \textit{DM-friendly graph store}, redesigning the traditional CSR format to alleviate the IOPS bottleneck by merging reads of indices and edge-lists/metadata. 
Second, to address the costly update propagation, \sysname proposes an \textit{adaptive update coordinator}. 
For dense iteration, it applies \textit{collaborative update} that takes cache affinity into consideration and runtime re-distributes update candidates to the node that is more suitable to handle them.
For sparse iteration, it switches to \textit{direct remote update}, eliminating synchronization overhead.
Third, to achieve fast and effective load balancing, \sysname employs a \textit{two-stage workload manager}. 
At startup, \sysname finds the coarse-grained partitioning can be fast while not compromising effectiveness.
At runtime, \sysname applies fine-grained re-scheduling to mitigate tail effect induced by hub-vertices.

We implement \sysname and evaluate it using various algorithms and billion-scale graphs. 
Our results reveal that, compared with the state-of-the-art graph processing system on DM, 
\sysname can elastically and independently scale up compute and memory resources, delivering up to 4.9$\times$ better performance and accommodating graphs with ever-increasing sizes;
meanwhile, it effectively tames the compute-side cache demands by up to 18.9$\times$, positioning itself as a DM-ready solution in practice.
Code will be open-sourced.

\section{Background}

\subsection{Graph Processing}\label{sec:bg:graph}

\begin{figure}[tbp]
  \centering
  \setlength{\abovecaptionskip}{5pt}
  \includegraphics[width=0.48\textwidth]{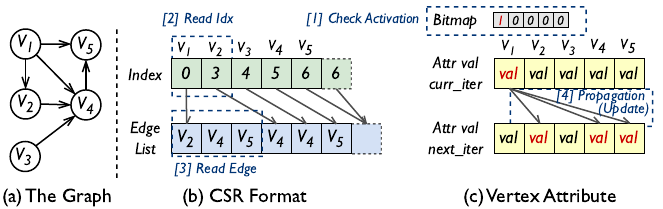}
  \caption{Data structures and workflow in graph processing.}
  \label{FIG_GRAPH_PROC}
\end{figure}

Graphs are commonly represented as $G = (V, E)$, where $V$ denotes the set of vertices, and $E \subseteq V \times V$ is the set of edges. An edge $(u,v)$ connects two vertices, directed from the \textit{source} vertex $u$ to the \textit{destination} vertex $v$, where $u$ is the \textit{in-neighbor} of $v$, and $v$ is the \textit{out-neighbor} of $u$. For space efficiency, graph data is typically stored in the compressed sparse row (CSR) format (Figure~\ref{FIG_GRAPH_PROC}(b)), which consists of an \textit{edge list array} that stores the out-neighbors for all vertices in a contiguous manner, and an \textit{index array} that records the starting position of each vertex's edge list. Thus, to retrieve the edge list of a vertex \textit{v}, the corresponding indices, $idx_{v}$ and $idx_{v+1}$, are read from the index array, and then the associated edge list located between $idx_{v}$ and $idx_{v+1}$ in the edge data can be accessed.

Based on the above-mentioned graph data structure, many \textit{graph algorithms} have been proposed to extract meaningful information. 
These algorithms typically maintain two vertex attribute arrays to hold vertex states/values for the current and next iterations (Figure~\ref{FIG_GRAPH_PROC}(c)). 
In each iteration, vertices that are \textit{active} according to the algorithm’s rule propagate their attribute values along outgoing edges to update their neighbors.
As shown in Figure~\ref{FIG_GRAPH_PROC}, a typical iteration involves four steps: (1) checking vertex activation, (2\&3) accessing corresponding indices and edge lists, and (4) propagating attribute values to neighbors. 
The algorithm ends when no vertices remain active or a predefined convergence condition is met. 
We refer to this whole procedure as \textit{graph processing} (or graph analytics). 
Notably, a vertex's neighbors are often scattered across the graph's data layout, and a destination vertex may be updated by multiple sources. 
This leads to highly random and contended memory accesses, posing a fundamental challenge for graph processing.

\subsection{Disaggregated Memory (DM)}~\label{sec:bg:dm}

\begin{figure}[thbp]
  \centering
  \setlength{\abovecaptionskip}{5pt}
  \includegraphics[width=0.4\textwidth]{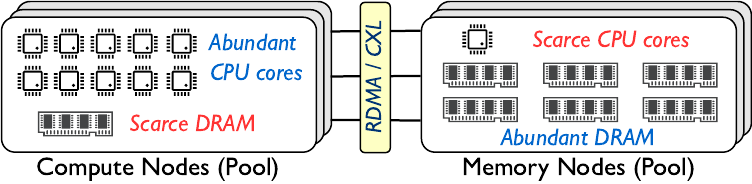}
  \caption{The Disaggregated Memory Architecture.}
  \label{FIG_DM_ARCH}
\end{figure}

The \textit{disaggregated memory (DM)} architecture has recently been proposed for cloud infrastructure~\cite{legoos_osdi,zombieland_eurosys_dm} to address the resource inefficiency in traditional data centers built on monolithic servers, which couple different resource components together~\cite{dm_moti_google,dm_moti_alibaba_1,dm_moti_alibaba_2}.
As shown in Figure~\ref{FIG_DM_ARCH}, DM decouples compute and memory resources into separate compute and memory pools. 
The compute pool consists of \textit{compute nodes} (CNs) with abundant CPU cores but scarce DRAM as local cache (e.g., 1--2 GB)~\cite{smart_osdi,sherman_sigmod}. 
The memory pool, on the other hand, comprises \textit{memory nodes} (MNs) with adequate DRAM yet weak computation power (e.g., 1--2 CPU cores)~\cite{aceso_sosp,dex_vldb,sherman_sigmod}.
The two pools are connected by advanced interconnect techniques with ever-increasing speed, such as remote direct memory access (RDMA)~\cite{rdma2023manual} or compute express link (CXL)~\cite{cxl2023manual}. 
By allocating different resources from the pools on-demand and independently, DM improves resource utilization and has become an important paradigm in industry~\cite{PolarDB-MP_sigmod24,GaussDB_vldb24}.
Notably, virtualization cannot fully address resource under-utilization: once a server's CPU cores are all rented, its remaining memory cannot be leased~\cite{li2023pond}.

Following prior works~\cite{kona_asplos21,smart_osdi,sherman_sigmod,dex_vldb,ditto_sosp,chime_sosp,aceso_sosp}, this paper focuses on \textit{RDMA-based DM}, as CXL 3.0 devices that enable memory pooling and sharing are not yet available.
RDMA offers RDMA verbs for developers to use, which include both \textit{one-sided verbs} (\texttt{READ}/\texttt{WRITE}/\texttt{CAS}/\texttt{FAA}) and \textit{two-sided verbs} (\texttt{SEND}/\texttt{RECV}). 
One-sided verbs enable direct remote memory access without involving the remote CPU, whereas two-sided verbs require remote CPU participation and can support more complex operations, such as memory management~\cite{fusee_fast,ditto_sosp,sherman_sigmod} and selective offloading tasks~\cite{dex_vldb,aceso_sosp,Dinomo_vldb,FG_sigmod}.

\section{Observation and Motivation}\label{sec:ob_moti}

\subsection{Graph Processing: Distributed vs. DM}
~\label{sec:motivate_graph_on_DM}
\subsubsection{\textbf{Issues of Distributed Graph Processing}}
Distributed graph processing systems have limitations in \textit{resource utilization}, particularly in terms of \textit{resource over-provisioning} and \textit{vertex replication}.

\textbf{Resource Over-Provisioning.}
Existing distributed graph processing systems are built on monolithic servers, where CPU and DRAM are tightly coupled, scaled at a fixed ratio, and cannot be shared across physical servers.
Such coupling is a well-known source of resource under-utilization in data centers~\cite{dm_moti_google,dm_moti_alibaba_1,dm_moti_alibaba_2}.
Tenants often need to over-provision servers to satisfy peak demand in either computation power or memory capacity.
Specifically, first, users may allocate multiple servers to achieve higher performance, while much of the attached memory remains under-utilized.
Second, large graphs that cannot fit in a single server require multiple servers for memory capacity, even though the coupled CPU resources may be under-utilized, e.g., for sparse queries.

\begin{figure}[tbp]
  \centering
  \setlength{\abovecaptionskip}{5pt}
  \includegraphics[width=0.5\textwidth]{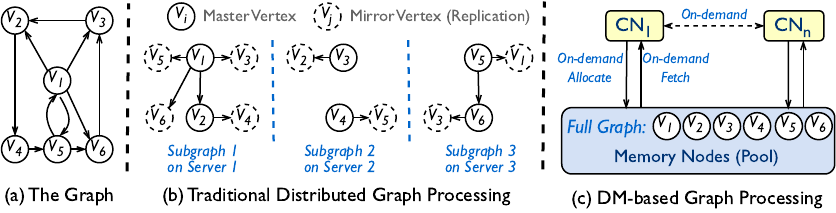}
  \caption{An example of vertex replication in distributed graph processing systems. DM-based graph processing places the graph on the large shared-accessible memory pool.}
  \label{FIG_REPLICA}
\end{figure}

\textbf{Vertex Replication.}
Moreover, we identify vertex replication as another issue in distributed graph processing.
Distributed graph processing systems typically partition the graph into multiple subgraphs and assign them to be held and processed by different servers.
Figure~\ref{FIG_REPLICA}(b)\footnote{Graph partitioning methods are generally categorized into edge-cut and vertex-cut. 
We use edge-cut partitioning as an example here, since it is widely used and has been shown to perform well in practice~\cite{partition_study_2018}.
Vertex-cut partitioning also naturally introduces vertex replication and thus does not change our motivation.}
shows how vertices are replicated across servers in distributed graph processing systems.
Vertices are initially assigned to different subgraphs (servers), and each server keeps the master copy of its assigned vertices~\cite{powergraph_2012,powerlyra_eurosys,gemini_2016,gluon_2018}.
The edges associated with a master vertex are also assigned to the same subgraph (server).
Once the graph is distributed across servers, a server may need to process edges that touch vertices whose master copies reside on other servers.
Existing systems therefore create local replicas of such vertices, called \textit{mirror vertices}, to support processing on partitioned subgraphs.
At the end of each iteration, mirrors synchronize with their master copies.
This phenomenon incurs non-negligible memory overhead, because finding a balanced partition with few cross-subgraph edges is NP-hard~\cite{NE_partition}, making replicas hard to avoid, while each replica also consumes extra memory for index and data of vertices~\cite{sigmod_ldbc_social_net}.
For example, on 4 servers with vertex replication, Gemini~\cite{gemini_2016} uses $2.7\times$ and $2.0\times$ more memory for vertices than on a single server for \textit{twitter-2010} and \textit{clueweb12} graphs, respectively.

\subsubsection{\textbf{Advantages of DM-based Graph Processing}}
DM-based graph processing systems can resolve the aforementioned issues.
First, for resource over-provisioning, DM enables on-demand allocation of compute and memory resources beyond the capacity of a single machine, better matching varying workload demands in a cost-efficient manner.
Second, DM can natively mitigate vertex replication, whose root cause is the isolated and capacity-limited local memory of monolithic servers. 
In distributed graph processing systems built on monolithic servers, graph data must be physically partitioned across servers, and replicas are introduced at partition boundaries to enable distributed execution. 
By contrast, DM provides a large shared memory pool directly accessible by multiple CNs. 
The graph can thus be stored in DM with a unified view, allowing multiple CNs to access a single copy of vertex data on demand (Figure~\ref{FIG_REPLICA}(c)) rather than maintaining replicas across servers. 
It in turn reduces the need for vertex replication and the associated memory overhead.

\subsection{Progress \& Limitations of Existing Systems}
~\label{sec:existing_work}
Recently, FAM-Graph~\cite{famgraph} and Fargraph~\cite{fargraph} made early attempts at graph processing on DM. 
Both systems offload edge data to RDMA-attached remote memory.
FAM-Graph follows the design of single-machine in-memory graph processing systems~\cite{ligra,galois}, operating at a fine-grained per-vertex level. 
In contrast, Fargraph is built on a storage-based graph processing system~\cite{gridgraph_2015}, benefiting from coarse-grained sequential access but failing to exploit DM’s fine-grained on-demand access capability.

Unfortunately, we found that existing works, both FAM-Graph and Fargraph, are \textit{severely impractical considering system scalability and compute-side cache demands}.
Consequently, they fall short of meeting the criteria and realizing the benefits of the memory-disaggregated architecture.

\begin{figure}[tbp]
  \centering
  \setlength{\abovecaptionskip}{5pt}
  \includegraphics[width=0.48\textwidth]{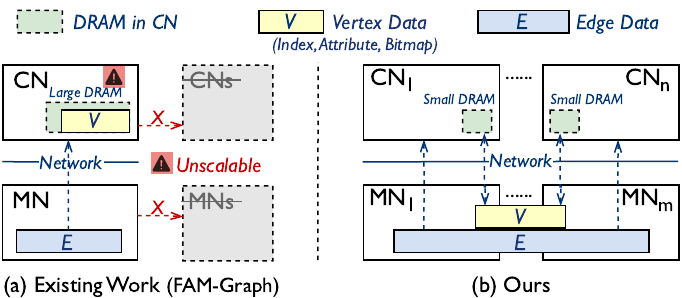}
  \caption{Different system architectures and data layouts for graph processing systems on DM. \textnormal{Existing works are unscalable and over-demanding of compute-side cache.}}
  \label{FIG_DIFF_ARCH}
\end{figure}

\textbf{Impracticality 1: \textit{Unscalable system.}} 
Notably, both FAM-Graph and Fargraph are restricted to a single CN and a single MN (Figure~\ref{FIG_DIFF_ARCH}(a)), unable to scale to multiple CNs or MNs. 
Such restrictive architecture is unacceptable.
First, it fails to handle large graphs which can easily exceed the memory capacity of a single MN. 
Second, it confines computation to the CPU cores of a single CN, resulting in long execution times for large graphs.
These constraints contradict the core objective of DM: \textit{enabling elastic resource provisioning to support diverse datasets and workload requirements}.
Moreover, they have not answered the questions of how to place graph data across MNs, how to distribute workloads among CNs, and how to handle concurrent updates from CNs to MNs. 
Unlike multiple threads within a single CN, accesses from CNs to MNs are both costly and lack native cache coherence, making this setting more challenging.

\textbf{Impracticality 2: \textit{Violating DM convention to over-demand compute-side cache.}}
To the best of our knowledge, DM systems assume only limited memory capacity (e.g., 1-2 GB) in CNs as local cache, thereby improving resource utilization by decoupling allocation of memory from computation resources. 
However, both FAM-Graph and Fargraph require substantial amounts of memory in each CN to hold the entire CSR index and all vertex attributes, which is impractical and misaligned with the DM architecture.
As reported in FAM-Graph, processing the \textit{clueweb12} dataset requires 22.5~GB of cache per CN~\cite{famgraph}, which is over an order of magnitude higher than typical DM configurations. 
Crucially, this is not a fixed overhead: the required compute-side cache scales linearly with both the number of vertices and the number of CNs, i.e., $O(|V|*\#CN)$.
As a result, processing larger graphs or applying more CNs proportionally increases the total compute-side memory usage.
This scaling re-couples memory allocation with CNs and can make aggregate CN memory usage rival or exceed MN usage, violating the goal of disaggregating memory from compute.

\subsection{Challenges}~\label{sec:motivation_challenge}
Based on the impracticalities of existing works, our goal is to build a graph processing system on DM that can \textit{\uline{scale to multiple CNs and MNs, require limited compute-side cache, and achieve high performance.}}
Figure~\ref{FIG_DIFF_ARCH}(b) shows our target system architecture and data layout.
All data in graph processing, including both edge data and full vertex data (indices, current- and next-iteration vertex attributes), reside in the DM pool made up by multiple MNs. The DM pool is viewed as a global address space shared by multiple CNs, addressed via a 16-bit MN ID and 48-bit in-MN offset, as prior works~\cite{sherman_sigmod,smart_osdi}. 
Each CN uses only a small DRAM (smaller than vertex data size) as explicitly managed cache to run large graphs.

Since this is still an unexplored problem, as a first step, we draw inspiration from traditional distributed graph systems and find that \textit{\uline{the chunk-based partitioning encourages a sensible approach for scaling graph processing to multiple nodes with constrained compute-side cache.}}
Typically, distributed graph systems~\cite{powergraph_2012,powerlyra_eurosys,gemini_2016,gluon_2018} partition vertices across servers, and each server executes the processing logic for its assigned vertices.
Gemini~\cite{gemini_2016} further exploits the empirical locality in large real-world graphs: neighboring vertices tend to reside near to each other.
Considering this property, Gemini introduces chunk-based partitioning, which divides the vertex set into $p$ contiguous, disjoint chunks, each assigned to one server. 
This layout offers good computation locality and keeps lots of update propagation within each server.
Based on the locality preserved in chunk-based partitioning, a sensible approach for graph processing system to scale to multiple nodes with constrained compute-side cache is: 
\textit{\uline{each CN caches only the corresponding chunk's next-iteration vertex attributes, which is randomly touched as shown in Figure~\ref{FIG_GRAPH_PROC}(c)}}. 
This approach offers three benefits: 
\begin{itemize}[leftmargin=*,topsep=0pt,parsep=\parskip,itemsep=0pt]
    \item A high fraction of random neighborhood updates can be absorbed by the CN cache thanks to locality in chunk, reducing remote accesses to MNs.
    \item Since it demands the CNs' aggregate memory to be compatible for a single attribute array, the system can process large graphs under tight CN cache and always scale to handle larger graphs by adding CNs or MNs as needed. For example, with 4 CNs, we require about 1~GB cache per CN for the largest \textit{clueweb12} dataset, an order of magnitude less than FAM-Graph, while aligning with typical DM architectures.
    \item Different CNs cache disjoint subsets of vertex attributes, simplifying cache-coherence management.
\end{itemize}

Built on the practical DM architecture and data layout for graph processing described above, several challenges still impede achieving high performance.

\textbf{Challenge 1: \textit{Retrieving graph from DM exhausts network.}}
Placing graph data on MNs forces CNs with limited local cache to perform costly remote memory accesses. 
The overhead of remote accesses becomes more severe considering the nature of graph workloads, whose access patterns are notoriously random, dependent, and fine-grained. 
\textit{\uline{(1) Randomness.}}
The neighbors of a vertex are typically scattered randomly in the graph file. Thus, graph processing often involves accessing arbitrary memory locations in an irregular pattern.
\textit{\uline{(2) Dependency.}}
Retrieving edges of a vertex requires two dependent high-latency remote accesses: one to fetch the index for the address and length of the edge list and a subsequent one for the edge data. Such dependency doubles latency for each vertex traversal.
\textit{\uline{(3) Fine granularity.}}
Specifically, graph data elements like vertex attributes and index entries are typically small, just a few bytes. 
Moreover, edge lists of most vertices are short due to the power-law degree distributions common in real-world graphs~\cite{powergraph_2012}.

\begin{figure}[tbp]
  \centering
  \begin{minipage}[t]{0.49\linewidth}
    \centering
    \setlength{\abovecaptionskip}{5pt}
    \includegraphics[width=\textwidth]{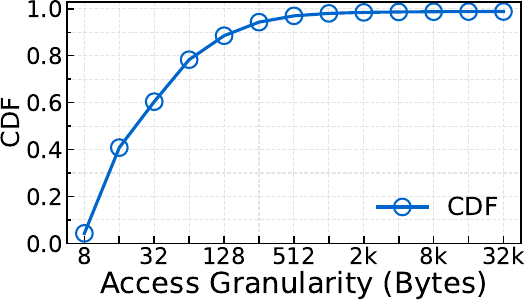}
    \caption{CDF of access granularity for graph retrieval during BFS on \textit{twitter-2010}}
    \label{FIG_MOTI_RETRIEVAL_SIZE}
  \end{minipage}
  \hspace{0em} 
  \begin{minipage}[t]{0.49\linewidth}
    \centering
    \setlength{\abovecaptionskip}{5pt}
    \includegraphics[width=\textwidth]{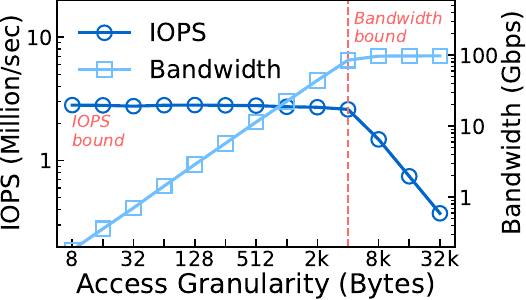}
    \caption{IOPS and bandwidth of \texttt{RDMA\_READ}~\cite{linux_rdma_perftest} for different access granularity}
    \label{FIG_MOTI_RETRIEVAL_IOPS}
  \end{minipage}
\end{figure}

Collectively, these characteristics lead to massive small remote accesses, which clashes with RDMA’s limited IOPS capacity. 
Figure~\ref{FIG_MOTI_RETRIEVAL_SIZE} summarizes the access granularity during a graph traversal after merging contiguous accesses (e.g., two adjacent required edge lists are fetched in one operation). 
Over 94\% of accesses are not larger than 256~B.
However, as Figure~\ref{FIG_MOTI_RETRIEVAL_IOPS} shows (measured by \texttt{perftest}~\cite{linux_rdma_perftest}), RDMA NICs are IOPS-bound for such small accesses (e.g., <4~KB), where protocol overheads dominate performance.

\textbf{Challenge 2: \textit{Remote updates to DM are unavoidable yet costly.}}
Due to real-world graphs' skewed degree distributions and cross-cutting connectivity, no graph partitioning method (including chunk-based method) can always yield perfectly isolated and size-balanced sub-graphs~\cite{powerlyra_eurosys,NE_partition}. 
As a result, a substantial fraction of edges inevitably span partitions. 
Consequently, when a CN processes such edges whose destination vertices lie outside its cached subset, update propagation along such edges requires a CN to perform remote update to corresponding vertex attribute on DM.

Unfortunately, remote updates are far more costly than local ones. 
Since multiple vertices can connect to the same vertex, multiple updates towards the same vertex attribute can be generated from multiple CNs.
To ensure correctness, DM systems typically use RDMA atomic operations (\texttt{RDMA\_CAS} and \texttt{RDMA\_FAA}) to enable concurrent updates from CNs~\cite{sherman_sigmod, chime_sosp}.
However, the remote updates based on RDMA atomic operations are costly for two reasons. 
First, the RDMA atomics are fine-grained, which can contend for limited network IOPS.
Second, they risk failures and retries by CNs' concurrent updates, further pressuring the network. 
Table~\ref{TAB_MOTI_REMOTE_UPDT} quantifies the impact under a full-activation iteration: despite chunk-based partitioning, only a small fraction of edges require remote updates (1.83--54.1\% for real-world graphs), yet they induce a substantial slowdown compared to an idealized all-local baseline (19.4--39.6$\times$).

Furthermore, in real graph processing workloads, update volumes incurred by different vertices or across different iterations are highly variable, precluding one-size-fits-all solutions.

\begin{table}[thbp]
  \centering
  \setlength{\abovecaptionskip}{3pt}
  \caption{Ratio of direct remote updates and corresponding slowdown rate (relative to local-only updates) using chunk-based partitioning in a fully-activated iteration (4 CNs).}
  \label{TAB_MOTI_REMOTE_UPDT}
  \begin{tabular}{ccccc}
      \toprule
      \textit{Dataset} & TW & UK & R29 & CW\\
      \midrule
      \textit{Ratio} & $54.1\%$ & $1.83\%$ & $72.0\%$ & $6.04\%$\\
      \textit{Slowdown} & $39.6\times$ & $19.4\times$ & $28.4\times$ & $29.7\times$\\
      \bottomrule
  \end{tabular}
\end{table}

\begin{figure}[tbp]
  \centering
  \begin{minipage}[t]{0.432\linewidth}
    \centering
    \setlength{\abovecaptionskip}{5pt}
    \includegraphics[width=\textwidth]{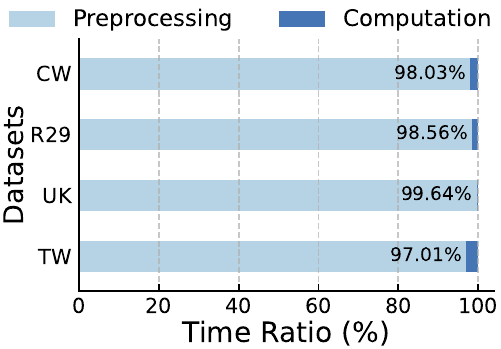}
    \caption{Preprocessing time ratio in Gemini (4 servers, PageRank): \textit{existing methods have long startup time}.}
    \label{FIG_MOTI_GEMINI_PREPROC}
  \end{minipage}
  \hspace{0em} 
  \begin{minipage}[t]{0.54\linewidth}
    \centering
    \setlength{\abovecaptionskip}{5pt}
    \includegraphics[width=\textwidth]{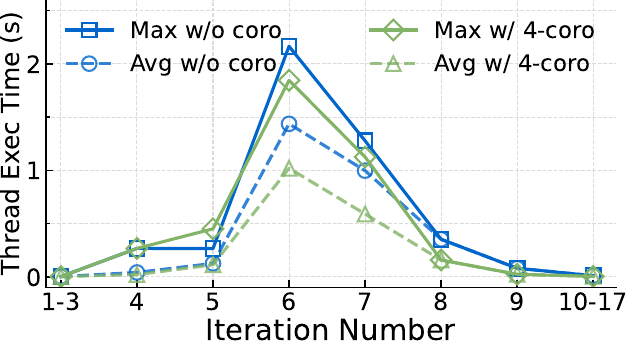}
    \caption{Execution time across threads in baseline DM graph system (1 CN; BFS; TW): \textit{both w/o and w/ coroutines, the maximum thread execution time is significantly longer than the average}.}
    \label{FIG_MOTI_RS}
  \end{minipage}
\end{figure}

\textbf{Challenge 3: \textit{DM urges fast and effective load balancing.}}
Although load balancing is an essential and well-studied topic for traditional parallel systems, graph processing on DM presents unique challenges, making existing methods ineffective.
\textit{\uline{(1) Inter-server load balancing}} typically relies on graph partitioning during the startup phase to build balanced subgraphs assigned to different servers. 
Unfortunately, as shown in Figure~\ref{FIG_MOTI_GEMINI_PREPROC}, state-of-the-art systems like Gemini~\cite{gemini_2016} require prohibitively long preprocessing time, greatly surpassing end-to-end gains.
This slow startup is especially problematic for DM, where systems may adaptively adjust compute resources to optimize resource efficiency, necessitating repeated system reconfiguration.

\textit{\uline{(2) Intra-server load balancing}} is further complicated by the high latency of remote memory accesses on DM. 
We first build a baseline system using the intra-server workload manager from a state-of-the-art design~\cite{gemini_2016}: 
each thread is initially assigned a set of vertices to process and, once finished, steals additional vertices from others.
As shown in Figure~\ref{FIG_MOTI_RS}, execution times vary significantly: the slowest thread becomes a clear outlier, bottlenecking overall progress.
Introducing coroutines within each thread (details in §~\ref{sec:design:optimization:coroutine}) hides some RDMA latency, but the tail effect persists. 
This stems from the interaction between hub-vertices and high-latency RDMA operations.
Real-world graphs with skewed degree distributions naturally contain hub-vertices that connect to an unusually large number of other vertices.
Processing them triggers numerous high-latency RDMA operations, leading to pronounced tail latency.
Coroutines offer limited relief: while they expose asynchronous execution, they can also concentrate multiple hub vertices in the same thread.

\section{\sysname Design}
\subsection{Overview}
\begin{figure}[thbp]
  \centering
  \setlength{\abovecaptionskip}{5pt}
  \includegraphics[width=0.45\textwidth]{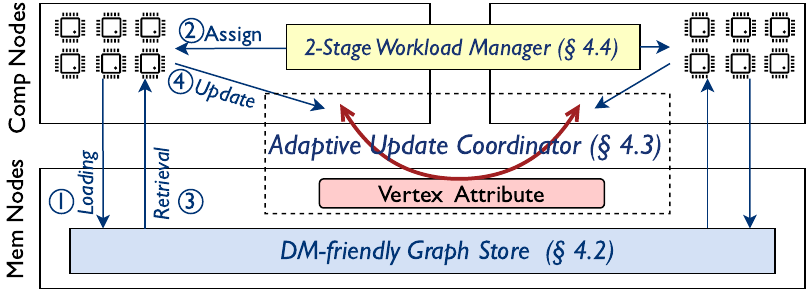}
  \caption{The Overview of \sysname.}
  \label{FIG_OVERVIEW}
\end{figure}

We propose \sysname, the first practical graph processing system on DM with system scalability, cache efficiency, and high performance. 
Figure~\ref{FIG_OVERVIEW} shows the overview of \sysname. 
\sysname decomposes graph processing workflow into four key steps and its designs comprehensively cover all of them: loading graph to DM(\ding{182}), workload assignment (\ding{183}), graph retrieval (\ding{184}), and update propagation (\ding{185}).

\sysname incorporates three designs to address the challenges outlined earlier.
First, for \textit{Challenge 1}, \sysname presents the \textit{DM-friendly graph store} in §~\ref{sec:design:graph_store} to enable efficient placement and retrieval of graph data on DM. 
The storage format also forms the basis for subsequent components.
Second, to resolve the complexities of concurrent remote updates from multiple CNs (\textit{Challenge 2}), \sysname proposes an \textit{adaptive update coordinator} in §~\ref{sec:design:coordination}.
Third, to tackle \textit{Challenge 3}, \sysname employs a \textit{two-stage workload manager} in §~\ref{sec:design:workload_manager} to achieve fast and effective load balancing both across and within CNs.
Finally, §~\ref{sec:design:optimization} presents optimizations and discussion.

\subsection{DM-friendly Graph Store}\label{sec:design:graph_store}

The storage format fundamentally determines the efficiency of storing and accessing graph data on DM.
Figure~\ref{FIG_GRAPH_STORE} shows \sysname's DM-friendly graph store. 
(1) \sysname stores vertex data (indices and attributes) only once by \textit{a single array} on DM, without replication in each CN or MN.
(2) \sysname redesigns the CSR format with an adaptive index scheme (§~\ref{sec:design:adaptive_index}) to improve retrieval efficiency and enable multi-node coordination.
(3) Moreover, two optimizations (§~\ref{sec:design:read_opt}) are applied to enhance graph retrieval on DM.
(4) Lastly, we detail the process and layout of loading graph to DM in §~\ref{sec:design:load_graph}.

\subsubsection{\textbf{Adaptive Index Scheme}}~\label{sec:design:adaptive_index}
As revealed in \textit{Challenge 1}, retrieving graph data from DM incurs substantial overhead, as frequent small remote accesses stress the limited IOPS of RDMA networks.
\sysname addresses this issue via an adaptive index scheme that aligns graph data with hardware characteristics.

Our key insight is to alleviate the IOPS bottleneck by placing the indices and low-degree edge lists together, turning many fine-grained RDMA reads into fewer, larger transfers.
Specifically, since RDMA throughput for small requests is limited by IOPS ~\cite{guideline_rdma} rather than bandwidth, increasing the access size per request enables more data to be transferred at the same IOPS cost. 
Moreover, the power-law degree distribution in real-world graphs, while it naturally leads to many fine-grained accesses on abundant low-degree vertices, also creates an opportunity: the small edge lists of these vertices can fit within a single, RDMA-efficient index block.

\begin{figure}[tbp]
  \centering
  \setlength{\abovecaptionskip}{5pt}
  \includegraphics[width=0.48\textwidth]{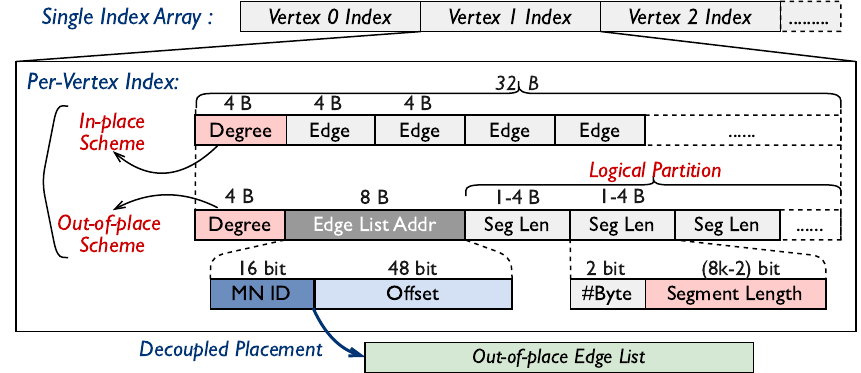}
  \caption{The DM-friendly Graph Store.}
  \label{FIG_GRAPH_STORE}
\end{figure}

Building on these observations and inspired by prior works on multi-level graph store~\cite{llama_icde_2015, graphone, terrace,LSMGraph_sigmod}, we redesign the classic CSR format for graph retrieval on DM. 
Instead of storing a single offset per vertex, we employ an expanded index structure with larger size (32B in this work, chosen to balance performance and memory efficiency; see Figure~\ref{FIG_EXP_IDX_SIZE}) that can either embed edges directly or store metadata for an out-of-place edge list. 
As illustrated in Figure~\ref{FIG_GRAPH_STORE}, \sysname chooses the in-place or out-of-place scheme for each vertex by comparing its recorded degree with the maximum number of edges that can fit in one index entry.
By slightly enlarging each RDMA request, \sysname obtains auxiliary information without additional network overhead, since the cost is dominated by the number of RDMA operations rather than their size.

\textbf{In-place edge index scheme.}
For vertices with few edges, \sysname stores the edge list directly in the index structure, identified as the in-place scheme. 
Edges are fetched via a single RDMA read rather than two as in traditional CSR-based graph systems.
Because real-world graphs are highly skewed, a large fraction of vertices have only a few neighbors and can be inlined with little memory overhead.
For example, in the \textit{twitter} graph~\cite{twitter}, the average degree is 39, while 65\% of vertices have fewer than 10 edges, making them natural candidates for the in-place scheme.

\textbf{Out-of-place edge index scheme.}
For high-degree vertices whose edge lists cannot fit within the index structure, \sysname adopts an out-of-place scheme: 
as shown in Figure~\ref{FIG_GRAPH_STORE}, each index entry stores the vertex degree, the global address of out-of-place edge list in the DM pool (16-bit MN ID + 48-bit offset in the MN), and per-segment lengths.
The out-of-place scheme includes three design concepts.
First, each edge list is logically partitioned into segments, where each segment contains edges whose destination vertices fall within a specific partition range (§~\ref{sec:design:partition}). 
The index explicitly records the length of each segment, allowing the system to identify and access the segment of edge list relevant to a given partition, which is crucial for efficient multi-node coordination (§~\ref{sec:design:coordination}).
Second, to reduce space overhead, we compress segment lengths using variable-length encoding: the first two bits indicate the number of bytes (1-4) used, and the remaining bits store the segment length. 
As Figure~\ref{FIG_EXP_SEG_LEN_BYTE} shows, most segments fit within a single byte, keeping the index structure compact. 
For extreme cases requiring more segments, the index can use more bytes per entry, incurring only limited space overhead (Figure~\ref{FIG_EXP_IDX_SIZE}).
Finally, the out-of-place edge lists are distributed independently of the vertex index across MNs to avoid hotspots in memory usage and network bandwidth caused by clustering of high-degree vertices.

\subsubsection{\textbf{Optimization: Merged and Batched Retrieval}}\label{sec:design:read_opt}
\sysname accelerates graph retrieval from DM using two optimizations: merging and doorbell batching~\cite{famgraph,DrTM_H}.
Merging combines multiple reads from contiguous memory into a single RDMA operation.
Doorbell batching allows the RNIC to fetch multiple requests by one DMA and signal completion only once. 
Together, these techniques improve network and PCIe utilization while reducing CPU overhead.

\begin{figure}[tbp]
  \centering
  \setlength{\abovecaptionskip}{5pt}
  \includegraphics[width=0.45\textwidth]{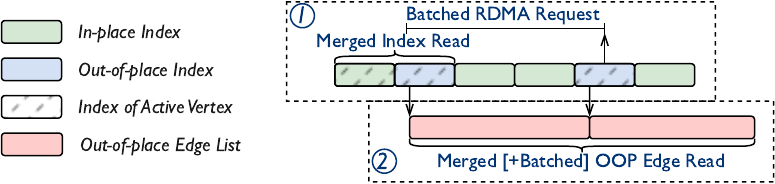}
  \caption{Merged and batched retrieval of graph data.}
  \label{FIG_READ_OPT}
\end{figure}

Figure~\ref{FIG_READ_OPT} shows \sysname's approach: each worker processes a batch of consecutive vertices at a time. 
Instead of operating vertex by vertex, \sysname adopts a batch-oriented workflow across all steps (activation check, index and edge-list retrieval) to reduce RDMA invocation overhead.
After batch-level activation checks, 
\sysname issues RDMA reads for indices of all active vertices, merging requests to contiguous regions and grouping them into doorbell batches.
Once the indices are fetched, \sysname identifies entries that reference out-of-place edge lists and similarly issues merged and batched RDMA reads for those edges. 
Notably, edge lists of consecutive vertices with the out-of-place scheme are placed contiguously during graph loading, ensuring batched reads either cover continuous regions or, at least, remain within the same MN. 
This locality is essential, as doorbell batching requires requests to be issued to the same RDMA queue pair (i.e., same MN).

\subsubsection{\textbf{Loading Graph to DM}}~\label{sec:design:load_graph}
\sysname leverages the large, globally addressable DM pool to simplify graph loading.
It loads and stores the edge list of each vertex in a continuous region on a single MN without scattering a vertex's edges across servers to construct subgraphs.
When building the graph store on DM from a graph in CSR format on a storage device, it proceeds as follows.

First, \sysname allocates a shared vertex index array in DM, which serves as the unified entry point for all edge data and is accessible by all CNs.
The array is range-partitioned across MNs: each MN stores a contiguous subset of vertex indices based on partitioning results.
\sysname allocates the vertex attribute array using the same policy.
Then, threads in CNs populate the graph store in parallel. 
Each thread loads a batch of vertices at a time and reads their edges from the CSR input.
For low-degree vertices, edges are embedded directly in the index structure. 
For high-degree vertices, \sysname allocates out-of-place memory space for the edge lists from the DM pool and records the addresses and segment lengths in the index structure.
To balance memory usage across MNs, the allocator for edges selects MNs in a coarse-grained round-robin manner. 
Out-of-place edge lists in the same batch are allocated in contiguous memory regions in one MN. 
After processing a batch, the thread writes the updated index entries to DM by a single RDMA operation.

\subsection{Adaptive Update Coordinator}\label{sec:design:coordination}

\begin{figure}[tbph]
  \centering
  \vspace{-0.3em}
  \setlength{\abovecaptionskip}{5pt}
  \includegraphics[width=0.48\textwidth]{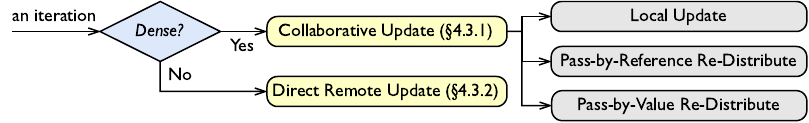}
  \caption{Different strategies in \sysname to perform update propagation in graph processing.}
  \label{FIG_ADA_PROC_WORKFLOW}
\end{figure}

For \textit{update propagation}, which encounters \textit{Challenge 2}, we propose an \textit{adaptive update coordinator} that adaptively chooses between different update strategies based on update density, cache affinity, and vertex characteristics. 
As shown in Figure~\ref{FIG_ADA_PROC_WORKFLOW}, when an iteration has dense updates (many active vertices), \sysname employs a novel \textit{Collaborative Update} scheme (§~\ref{sec:design:update:collaborative}), a key contribution of this work.
This scheme exploits cache affinity across CNs and MNs, and runtime redistributes update candidates that cannot be handled locally to nodes that can process them using local memory accesses.
The \textit{re-distribute (RD)} mechanism combines two complementary patterns, \textit{Pass-by-Value} and \textit{Pass-by-Reference}.
When an iteration has sparse updates (few active vertices), \sysname switches to a \textit{Direct Remote Update} scheme (§\ref{sec:design:update:direct_remote}), which simplifies execution and removes synchronization overhead.

\subsubsection{\textbf{Collaborative Update}}\label{sec:design:update:collaborative}
\sysname introduces a \textit{Collaborative Update} scheme for iterations with dense updates, which dominate the end-to-end execution time of graph processing workloads. 
Notably, as analyzed in §~\ref{sec:motivation_challenge}, chunk-based partitioning encourages an effective compute-side caching strategy: each CN caches only the attributes of vertices in its assigned chunk. 
This strategy ensures a high fraction of updates can hit the cache while requiring only limited compute-side cache, even for increasingly large graphs.
However, as revealed in \textit{Challenge 2}, 
performance bottlenecks persist due to costly remote memory operations when updates target vertices that are not cached.

We identify two properties of chunk-based caching that open new opportunities in the design space: 
(1) \textit{\uline{The attribute of each vertex is guaranteed to be cached at some CN and stored at some MN;}}
(2) \textit{\uline{The CN or MN holding the attribute of a given vertex can be quickly located via the partition offsets.}}

Therefore, our key insight is that we can pass uncached update candidates\footnote{An update candidate indicates the attribute of the active source vertex and the destination vertices (edges) to be updated.} to the CN or MN that already holds the corresponding vertex attributes, and let that node execute the update logic via local memory operations.
In other words, we move updates to where the data resides via large sequential accesses, rather than forcing the CN that initially generates the update candidates to execute the update logic through costly random remote memory operations.
Since the passing process can batch information about many update candidates into a few coarse-grained sequential \texttt{RDMA\_WRITE/READ}, 
its overhead is substantially lower than directly executing fine-grained update logic over the network.

\begin{figure}[tbp]
  \centering
  \setlength{\abovecaptionskip}{6pt}
  \includegraphics[width=0.48\textwidth]{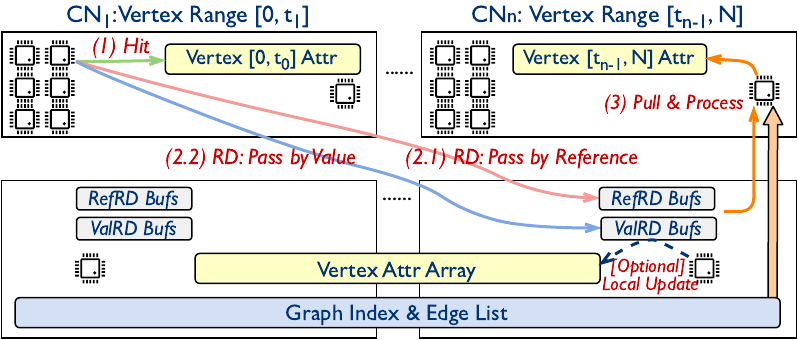}
  \caption{An example of \textit{Collaborative Update}: how to process the update candidates initially generated from $CN_1$.}
  \label{FIG_RD_WORKFLOW}
\end{figure}

As illustrated in Figure~\ref{FIG_RD_WORKFLOW}, consider a source vertex $v$ processed by $CN_1$, which belongs to the partitioned chunk $V_1 \subseteq V$. 
$CN_1$ loads and caches the attributes of vertices in $V_1$. 
When processing $v$'s edge list, if a destination vertex also resides in $V_1$, the update is performed directly on $CN_1$, i.e., \textbf{Local Update}. 
If the destination vertex belongs to another chunk, e.g., $V_n$, the update candidate is re-distributed to the responsible $CN_n$ or corresponding MN, which then executes the update locally, i.e., \textbf{Re-Distribute Update}. 
Re-Distribute (RD) includes two patterns: \textit{Pass-by-Reference Re-Distribute (RefRD)} and \textit{Pass-by-Value Re-Distribute (ValRD)}.

\begin{figure}[tbp]
  \setlength{\abovecaptionskip}{5pt}
  \subfigure[RD structure of passing by reference: at the granularity of an edge segment.] {
      \label{FIG_REF_RD}
      \centering
      \includegraphics[width=0.553\columnwidth]{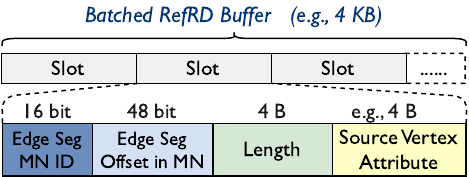}
  }
  \hspace{0.2em} 
  \subfigure[RD structure of passing by value: at the granularity of an edge] {
      \label{FIG_VAL_RD}
      \centering
      \includegraphics[width=0.38 \columnwidth]{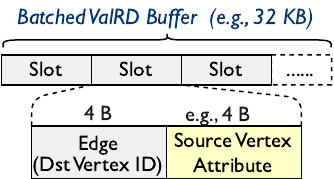}
  }
  \caption{Structure of Batched Re-Distribution (RD) Buffer}
  \label{FIG_RD_STRUCT}
\end{figure}

\textbf{Pass-by-Reference Redistribute.}
For vertices with long uncached edge-list segments, \sysname adopts pass-by-reference redistribution (\textit{RefRD}).
Instead of transferring the entire edge list, \textit{RefRD} transmits only metadata for each relevant segment, including the segment's starting address, length, and source vertex attribute, as illustrated in Figure~\ref{FIG_RD_STRUCT}(a). 
The pre-embedded segment lengths introduced in §~\ref{sec:design:adaptive_index} enable quick identification of the relevant segments for destination vertices within a given partition (chunk).
Segments whose destination vertices fall into different ranges are passed to different CNs.
\textit{RefRD} items are buffered at the sender CN and flushed to the corresponding MN either when the buffer is full or upon completion of the CN's workload.
Each CN uses a thread to pull the \textit{RefRD} buffer passed from other CNs. 
Upon receiving \textit{RefRD} items, the CN retrieves the corresponding long edge-list segment from DM and applies updates to the destination vertex attributes cached in its local memory.

\textbf{Pass-by-Value Redistribute.}
For vertices with short edge lists embedded in the index (e.g., degree $\leq 7$), as well as for long edge lists where only a small fraction of segments (e.g., segments with $\leq 7$ edges) fall outside the CN's partition, \sysname uses pass-by-value redistribution (\textit{ValRD}) instead of \textit{RefRD}.
As Figure~\ref{FIG_RD_STRUCT}(b) shows, each \textit{ValRD} item contains the source vertex attribute and the destination vertex ID. 
Similar to \textit{RefRD}, \textit{ValRD} items are also buffered, transferred, and processed in batches to eliminate fine-grained RDMA operations. 
Unlike \textit{RefRD}, which requires accessing out-of-place edge lists, \textit{ValRD} items already carry the information needed for update propagation. 
Therefore, these updates can be performed locally when the limited computation power of MNs is available and under low stress~\cite{dex_vldb,aceso_sosp,Dinomo_vldb}.
As confirmed by experiments, a single MN thread can handle the offloaded \textit{ValRD} items and perform updates locally with limited CPU usage.

Critically, the two RD patterns are complementary under the power-law degree distributions prevalent in real-world graphs, where most vertices have only a few edges while a small number of vertices have extremely many edges. 
Using only one pattern performs poorly, whereas combining both is essential for efficiency.
(1) A \textit{RefRD}-only design would generate a large number of references for low-degree vertices, forcing each receiver CN to issue costly fine-grained reads for every small segment.
(2) A \textit{ValRD}-only design is inefficient for high-degree vertices, as it must transfer large volumes of raw edge data, stressing network bandwidth and potentially saturating the MN's limited compute capacity.

At the end of each iteration that uses the collaborative update scheme with RD, CNs merge (evict) their local cached copies of vertex attributes with the attributes in the remote memory pool. 
The merging (eviction) is performed sequentially and without inter-CN contention, significantly improving over issuing fine-grained remote updates directly.

Notably, although RD relies on the partitioning result to determine the destination CN, and partitioning is embedded in the graph store (the out-of-place scheme), RD does not necessitate re-partitioning the graph or reloading the graph store whenever the number of CNs changes. 
Instead, the partitioning result and the graph loaded in DM can be reused across different CN counts. 
With fewer CNs, a graph store originally built for a larger number of partitions can be reused by aggregating multiple partitions.
With more CNs, the CNs can be organized into multiple groups that share the same graph store in DM.
For example, when scaling from 4 CNs to 8 CNs, the CNs can be organized into two 4-CN groups that reuse the same loaded graph.

\subsubsection{\textbf{Direct Remote Update}}\label{sec:design:update:direct_remote}
In many graph workloads, most iterations activate only a very small fraction of vertices (i.e., few updates).
For example, in a breadth-first search on the \textit{clueweb12} dataset, convergence occurs in about 500 iterations, yet in fewer than 50 iterations more than 0.01\% of vertices are activated for processing.
During such sparse iterations, CNs do not cache vertex attributes; instead, they perform updates directly via remote memory accesses for simplicity and efficiency. 
Since RDMA provides only limited atomic operations (\texttt{RDMA\_CAS}, \texttt{RDMA\_FAA}), updating a vertex attribute, such as writing a smaller value, requires multiple RDMA operations. 
Specifically, the CN first reads the current vertex attribute from remote memory and compares it locally; if an update is needed, it repeatedly issues \texttt{RDMA\_CAS} until the update succeeds or becomes unnecessary.

\subsection{Two-Stage Workload Manager}\label{sec:design:workload_manager}
To address \textit{Challenge 3}, \sysname employs a two-stage workload manager that operates across and within CNs.
The coarse-grained initial partitioning (§~\ref{sec:design:partition}) counteracts the prohibitive preprocessing overhead, 
while the fine-grained runtime re-scheduling (§~\ref{sec:design:RS}) alleviates the long-tail effect.

\subsubsection{\textbf{Coarse-Grained Initial Partitioning}}\label{sec:design:partition}
Generally, \sysname follows \textit{chunk-based partitioning}~\cite{gemini_2016} to achieve inter-CN load balancing: 
the vertex set is divided into subsets, and each subset of vertices is assigned to be processed by a CN.  
Formally, given a graph $G=(V, E)$, the vertex set $V = \{v_0, \dots, v_{n-1}\}$ is divided into $p$ disjoint, contiguous subsets (partitions) $V_0, V_1, \dots, V_{p-1}$ such that:

\begin{itemize}
\item $V = \bigcup_{i=0}^{p-1} V_i$ with $V_i \cap V_j = \emptyset$ ($i \neq j$)
\item $V_i = \{v_j \mid t_i \leq j < t_{i+1}\}$, for $0 = t_0 < \dots < t_p = n$
\end{itemize}

We apply the same objective function as Gemini~\cite{gemini_2016}, trying to balance the value of $8*(p-1)|V_i|+|E_i|$across partitions. $|V_i|$ is the vertex number and $|E_i|$ is the number of edges associated with $V_i$.

Our key observation is that vertex-level granularity in chunk-based partitioning is unnecessary and expensive when $|V|\gg p$. Existing methods traverse the full graph to gather per-vertex statistics, yet a graph often contains millions or billions of vertices while the partition count $p$ is only in the single or low double digits. The finest granularity marginally improves load balance but incurs a major partitioning cost.

To address this inefficiency, \sysname introduces the \textit{coarse-grained initial partitioning}.  
We first slice the vertex set into a fixed number (e.g., 1024) of contiguous \textit{tiny-chunks} and record lightweight metadata for each tiny-chunk, including vertex and edge counts.
Partitioning then manipulates only this metadata: \sysname assembles tiny-chunks into $p$ partitions and assigns them to CNs without revisiting the full edge list.  
The reusable tiny-chunks' metadata enables rapid re-partitioning when the DM system adjusts resources. 
To our knowledge, this is the first graph partitioning method that operates at the coarse granularity of tiny-chunks, rather than at the traditional vertex level.

\subsubsection{\textbf{Fine-Grained Runtime Re-Scheduling}}\label{sec:design:RS}
After initial (chunk-based) partitioning which divides the vertex set into multiple subsets and assigns each subset to a CN for processing, \sysname uses a fine-grained runtime approach within each CN to balance workload across threads and mitigate tail latency caused by hub-vertices.

\textit{(1) Per-thread chunking and work-stealing.}
Each CN further divides its assigned vertex subset into thread-local chunks, managed by per-thread progress managers.  
Threads process their local chunks first and employ work-stealing upon completion, reducing contention and improving utilization.
This mechanism was proved effective in Gemini~\cite{gemini_2016}, and \sysname adopts it as the basis for intra-CN load management.

\textit{(2) Runtime Re-Scheduling (RS).}
Previous steps assign workload only at the vertex granularity, leaving all edges of a few vertices to a single worker.
As revealed in \textit{Challenge 3}, this creates severe tail effects on DM: long edge lists trigger many high-latency network operations and stall overall progress. 
To address this issue, \sysname proposes \textit{Runtime Re-Scheduling (RS)}, a simple yet effective runtime mechanism that decomposes hubs into smaller segments.
When a worker encounters a vertex with high degree (e.g., >1024), \textit{RS} splits its edge list into segments and enqueues them into a per-thread \textit{RS buffer}, where segments await processing by (multiple) workers.  
In each thread, \sysname runs multiple coroutine workers (§~\ref{sec:design:optimization:coroutine}) that prioritize processing items from this buffer before fetching new vertices, allowing \textit{RS} to operate effectively within the thread and avoid synchronization overhead.
Our evaluation confirms that \textit{RS} effectively mitigates hub-vertex-induced stragglers and improves performance scalability as thread counts grow.

\subsection{Optimizations and Discussions}\label{sec:design:optimization}

\subsubsection{\textbf{Coroutine}}\label{sec:design:optimization:coroutine}
Coroutines are widely used in RDMA-based systems to hide the latency of RDMA operations~\cite{DrTM_H}.
We employ multiple (4 by default) coroutines per thread to execute graph processing tasks.
Each coroutine yields after issuing RDMA requests and resumes execution upon receiving request completions.
The asynchronous nature of coroutines overlaps computation and communication, enhancing overall performance.

\subsubsection{\textbf{Dual-mode Execution}}\label{sec:design:discussion:dual-mode}
Push-pull dual-mode execution is a popular optimization in graph processing systems~\cite{ligra,gemini_2016}, aiming to balance I/O volume and contention.
\sysname employs either push or pull mode adaptively depending on contention severity of different algorithms.
By default, it uses push mode for algorithms like breadth-first search (BFS).
For high-contention algorithms such as PageRank, where every edge is accessed in each iteration and the data transfer volume is similar in both modes, \sysname prefers to use pull mode.
In pull mode, \textit{ValRD} enables the MN to aggregate partial results produced by a compute-side worker.

Currently, \sysname does not perform mode selection in each iteration, but chooses a mode across iterations.
The selective use of pull mode is limited by two factors: limited experimental hardware and memory efficiency.
First, on our testbed with 100~Gbps RNICs, limited network bandwidth prevents us from aggressively trading more data transferred for less contention. 
Fortunately, the advent of RDMA and CXL with higher bandwidth (e.g., 800~Gbps and higher~\cite{D-RDMA_CIDR2022,Tassel_SIGCOMM2024}) can enable more flexible mode switching in the future. 
Second, supporting dual-mode for each query requires extra auxiliary in-memory structures for bidirectional access~\cite{FaaSGraph}.

\subsubsection{\textbf{Dynamic Graphs}}\label{sec:design:discussion:dynamic}
\sysname focuses on improving the performance of graph analytics queries over static graphs, as efficient processing of static graphs on DM is already a fundamental and non-trivial problem. 
Supporting dynamic graphs with edge insertions and deletions~\cite{terrace,Dynamic_Graph_sigmod25,radixgraph} is also an important direction, but is beyond the scope of this paper. 
However, \sysname preserves a natural path toward future dynamic support. 
In particular, \sysname's data layout and update-propagation mechanisms are amenable to graph updates. 
For edge insertions, the in-place index scheme provides a natural buffer for newly added edges on low-degree vertices. 
For vertices with out-of-place edge lists, a temporary out-of-place list can be maintained and periodically merged back. 
Edge deletions can be supported by marking removed entries with a tombstone value. 
Notably, newly inserted edges that have not been incorporated into partitioned segments can still be correctly handled with low overhead by using \textit{ValRD} for uncached update propagation.

\subsubsection{\textbf{When It Comes to CXL}}\label{sec:design:discussion:cxl}
Although this paper focuses on RDMA-based DM due to the limited availability of CXL devices, the proposed techniques are compatible and remain beneficial for CXL-based DM~\cite{Lupin,CXL-DSM,TrEnv,cxlmemsim_2023}. 
\textbf{For compatibility}, most of \sysname's memory accesses rely on one-sided verbs, which are also supported by CXL devices. 
And the \textit{ValRD} can also be pulled and executed by corresponding CN via coarse-grained sequential memory accesses. 
\textbf{For effectiveness}, first, the designs for reducing fine-grained accesses and hiding latency (§~\ref{sec:design:graph_store} and §~\ref{sec:design:coordination}) naturally carry over to CXL, as latency gaps between local and remote memory accesses persist on DM, and peak CXL memory bandwidth still requires accessing at sufficiently large granularity~\cite{cxl_mem_perf_vldb25}.
Second, our fast and effective workload manager (§~\ref{sec:design:workload_manager}) can benefit all multi-server graph processing systems, including those built on CXL-based DM.

\section{Evaluation}\label{sec:eval}

\begin{figure*}[ht]
  \begin{multicols}{2}
      \centering
      \setlength{\abovecaptionskip}{2pt}
      \includegraphics[width=0.48\textwidth]{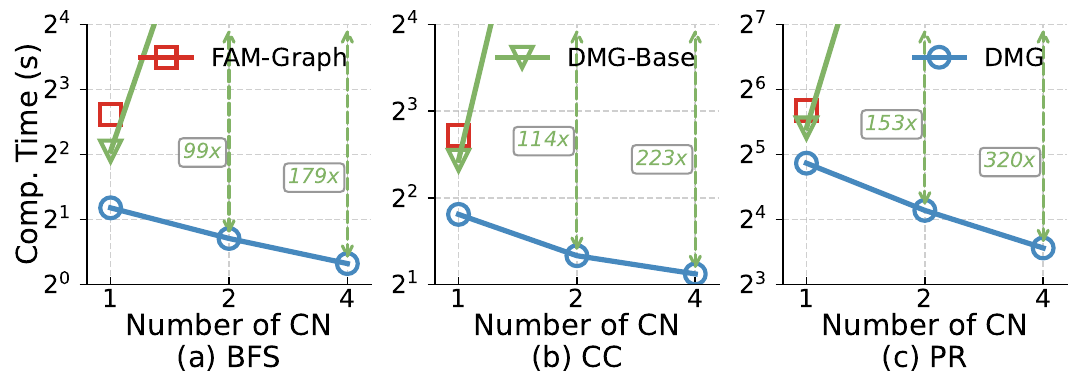}
      \caption{Computation time on \textit{twitter-2010}.}
      \label{FIG_PERF_COMP_TW}

      \setlength{\abovecaptionskip}{2pt}
      \includegraphics[width=0.48\textwidth]{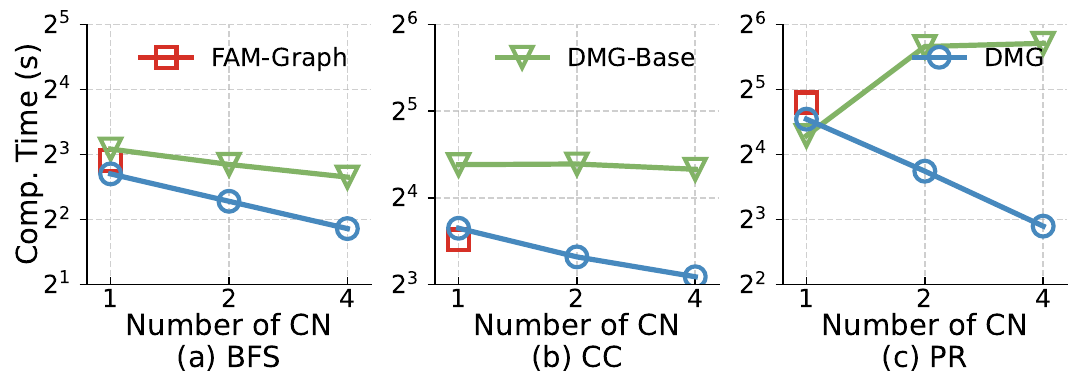}
      \caption{Computation time on \textit{uk-2007-05}.}
      \label{FIG_PERF_COMP_UK}
      
      \columnbreak 

      \setlength{\abovecaptionskip}{2pt}
      \includegraphics[width=0.48\textwidth]{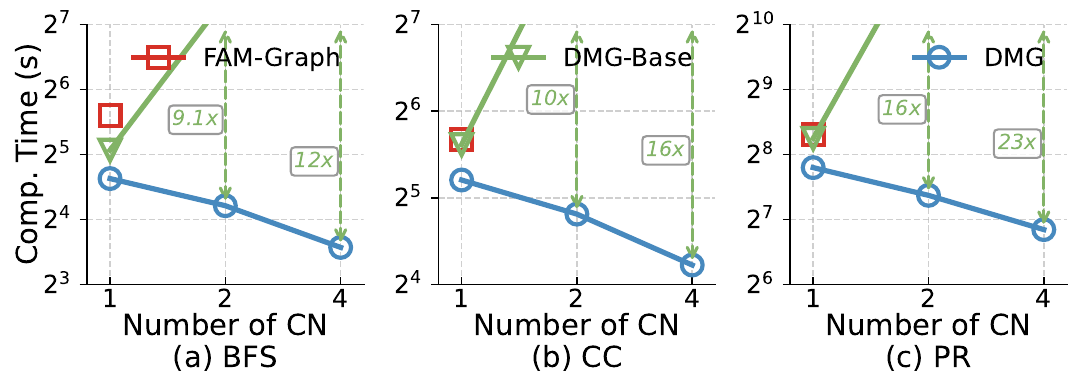} 
      \caption{Computation time on \textit{rmat-29}.}
      \label{FIG_PERF_COMP_R29}

      \setlength{\abovecaptionskip}{2pt}
      \includegraphics[width=0.48\textwidth]{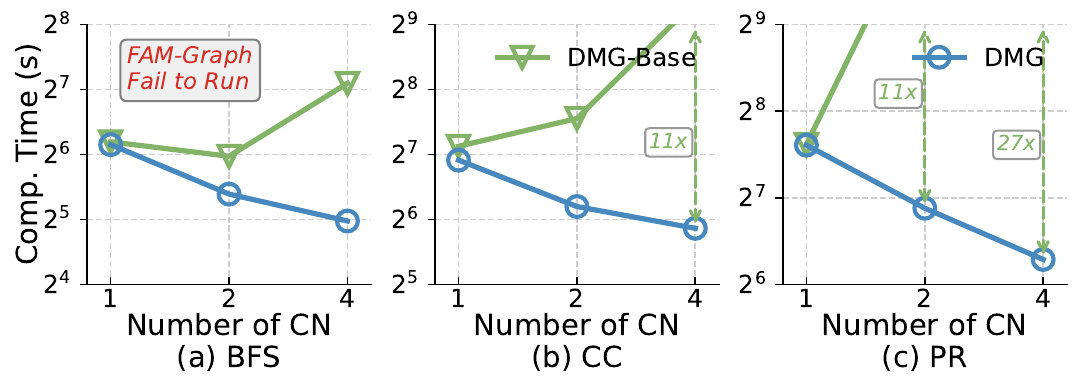} 
      \caption{Computation time on \textit{clueweb12}.}
      \label{FIG_PERF_COMP_CW}
  \end{multicols}
\vspace{-5pt}
\end{figure*}

\begin{figure}[tbp]
  \centering
  \setlength{\abovecaptionskip}{3pt}
  \includegraphics[width=0.48\textwidth]{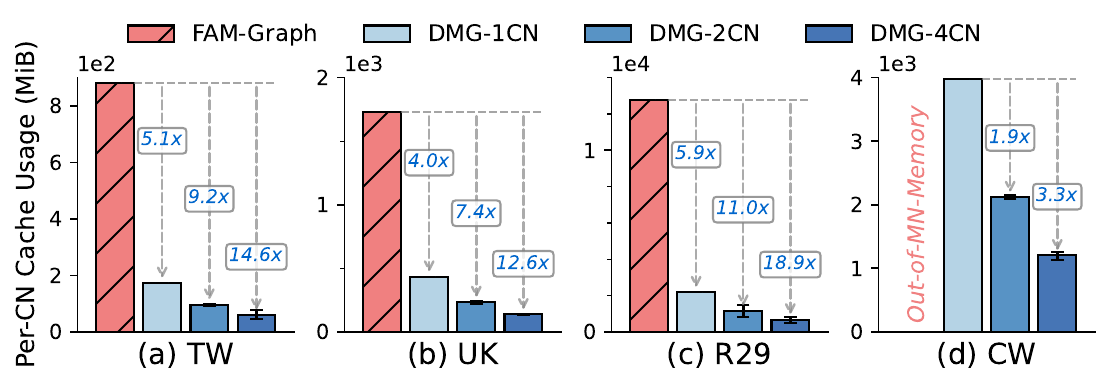}
  \caption{Cache demands in each CN.}
  \label{FIG_EXP_CN_MEM}
\end{figure}

\subsection{Experimental Setup}
\textbf{\textit{Testbed.}} 
We conduct experiments on 8 physical machines on the Utah cluster of CloudLab~\cite{cloudlab}.
Each machine is equipped with a 24-core EPYC 7402P CPU, 128GB DRAM, PCIe v4.0 NVMe SSD, and a 100Gbps Mellanox ConnectX-5 NIC.
The machines are interconnected via a 100Gbps switch. 
All machines run Ubuntu 22.04 with Linux kernel version 5.15.0. 
Each physical machine is configured to simulate either a CN or a MN.
The memory pool consists of 4 MNs, each with 128~GB of DRAM.
Each MN runs two threads for RPC serving and selective offloading, respectively, both of which have low CPU usage.
Unless specified, each CN uses 16 threads pinned to physical cores and up to 2~GB of DRAM.
Hugepages are applied to reduce address
translation cache misses in RDMA NICs.


\textbf{\textit{Datasets and Workloads.}}
We use four billion-scale graphs, listed in Table~\ref{tab:dataset}, for evaluation, including \textit{twitter-2010 (TW)} as a social network graph, \textit{rmat-29 (R29)} as a synthetic graph generated by R-MAT~\cite{dataset_rmat} with default parameters, \textit{uk-2007-05 (UK)} and \textit{clueweb12 (CW)}~\cite{dataset_webgraph} as web crawler graphs.
We evaluate three representative graph processing workloads: breadth-first search (BFS), connected components (CC), and PageRank (PR).
BFS starts from non-isolated vertices.
PR runs for 10 iterations with all vertices activated to obtain stable results.

\begin{table}[th]
  \centering
  \setlength{\abovecaptionskip}{5pt}
  \caption{Graph datasets used in evaluation.}
  \label{tab:dataset}
  \begin{tabular}{ccccc}
      \toprule
      Graph & $|V|$ & $|E|$ & Type\\
      \midrule
      \textit{twitter-2010 (TW)} & 42 M & 1.47 B & social network\\
      \textit{uk-2007-05 (UK)} & 106 M & 3.7 B & web\\
      \textit{rmat-29 (R29)} & 537 M & 8.6 B & synthetic\\
      \textit{clueweb12 (CW)} & 978 M & 42.6 B & web\\
      \bottomrule
    \end{tabular}
\end{table}

\textbf{\textit{Implementation and Comparisons.}}
We implement \sysname in about 11K lines of C++ code, and it will be open-sourced.
We evaluate \sysname primarily against FAM-Graph~\cite{famgraph}, a state-of-the-art graph processing system on disaggregated memory, 
and against \sysname-Base, a baseline that preserves the same system architecture and compute-side cache usage but disables all our proposed designs.
We omit comparisons with Fargraph~\cite{fargraph} for brevity, since it adopts a similar impractical system architecture to FAM-Graph but exhibits significantly lower performance due to its coarse-grained access pattern.
We also compare \sysname with Gemini~\cite{gemini_2016}, a state-of-the-art distributed graph processing system designed for monolithic servers, in §~\ref{sec:eval:distributed}.

\subsection{Overall Comparison}

This section presents an overall comparison of \sysname against FAM-Graph and \sysname-Base. 
Our goal is to evaluate whether \sysname can (1) avoid the unscalable architectures of prior systems, (2) eliminate their impractical cache demands, and (3) simultaneously deliver high performance on top of the scalable, cache-efficient DM architecture.
Specifically, Figure~\ref{FIG_PERF_COMP_TW}–\ref{FIG_PERF_COMP_CW} report the computation time across four datasets and three representative workloads. Figure~\ref{FIG_EXP_CN_MEM} shows the corresponding per-CN cache demands, which remain stable across workloads for a given system and dataset.

\textbf{(1) \sysname has a scalable and elastic DM architecture.}
\sysname eliminates the rigidity of FAM-Graph in both memory and compute. 
On the memory side, FAM-Graph supports only one MN and thus cannot handle graphs exceeding a single MN's capacity (e.g., \textit{clueweb12} in Figure~\ref{FIG_PERF_COMP_CW}). 
In contrast, \sysname aggregates multiple MNs into a large shared memory pool, enabling support for large graphs. 
On the compute side, FAM-Graph runs only on a single CN, whereas \sysname can operate efficiently with either a few CNs or many CNs while keeping memory-pool usage constant, thereby providing scalable, decoupled, and on-demand provisioning of compute and memory resources.

\textbf{(2) \sysname demands limited compute-side cache.}
As shown in Figure~\ref{FIG_EXP_CN_MEM}, \sysname dramatically reduces per-CN cache requirements compared to FAM-Graph. 
With one CN, \sysname consumes just $17$--$25\%$ of FAM-Graph’s cache footprint by placing all data in the MN pool and caching only essential slices. 
Moreover, adding more CNs further reduces per-CN cache to $30$--$34\%$. 
This property offers two benefits:
First, larger graphs can be supported under the same cache-size constraints by scaling out CNs.
Second, the CNs' aggregate memory remains a small fraction of the MNs' memory, avoiding replication overhead and preserving resource disaggregation.

\textbf{(3) \sysname achieves high performance.}
With a single CN, despite requiring less compute-side cache to keep the DM architecture practical, \sysname still achieves 1.76--2.71$\times$, 0.91--1.19$\times$, and 1.38--1.96$\times$ speedups over FAM-Graph on TW, UK, and R29 respectively. 
These gains stem from three designs: optimized data layout for faster graph retrieval, runtime rescheduling to mitigate tail effects, and coroutine-based asynchronous execution to overlap compute and communication. 

Scaling CNs further strengthens these advantages. 
When increasing from one to four CNs, \sysname accelerates end-to-end performance by $1.47$--$3.13\times$. 
Compared to the baseline \sysname-Base, \sysname delivers improvements of up to two orders of magnitude.
The large gap arises because hub-vertex tails and costly remote updates exacerbate each other in \sysname-Base but are simultaneously mitigated in \sysname. 
The only moderate case is the \textit{UK} dataset, where improvements remain at $1.7$--$8.0\times$, owing to its sparse cross-partition edges and less skewed degree distribution.

\subsection{Comparison with Traditional Distributed Graph Processing System}\label{sec:eval:distributed}

This section compares \sysname, a graph processing system built on DM, with Gemini~\cite{gemini_2016}, a state-of-the-art distributed graph processing system designed for monolithic servers. 
We evaluate three aspects: startup time, computation time, and resource efficiency.

\begin{figure}[tbp]
  \centering
  \setlength{\abovecaptionskip}{3pt}
    \includegraphics[width=0.45\textwidth]{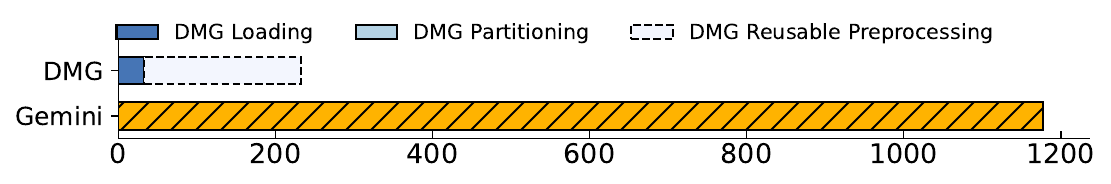}
  \caption{Startup time (s) of \sysname and Gemini for \textit{clueweb12} with 4 servers/CNs.}
  \label{FIG_EXP_GEMINI_STARTUP}
\end{figure}

To begin with, as shown in Figure~\ref{FIG_EXP_GEMINI_STARTUP}, \sysname substantially reduces graph startup time. 
For the billion-vertex \textit{clueweb12} graph, its startup time is only $2.8\%$ of Gemini's. 
Even when including the one-time preprocessing cost, whose results can be reused across executions, the total startup cost remains only $19.7\%$ of Gemini's.
This improvement comes from two factors. 
First, \sysname loads the graph directly into a unified large DM pool, stores each vertex's adjacency list contiguously, and avoids the costly physical partitioning and per-server data-structure construction required by traditional distributed systems. 
Second, its preprocessing results are reusable, eliminating repeated graph partitioning across executions.

\begin{figure}[tbp]
  \centering
  \setlength{\abovecaptionskip}{3pt}
    \includegraphics[width=0.48\textwidth]{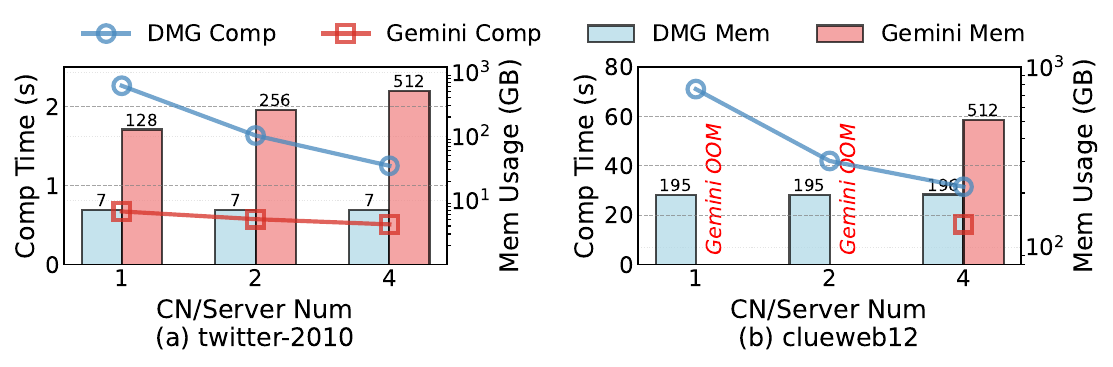}
  \caption{Memory usage and BFS computation time of \sysname and Gemini.}
  \label{FIG_EXP_GEMINI_COMP_MEM}
\end{figure}

After the graph is loaded, Figure~\ref{FIG_EXP_GEMINI_COMP_MEM} compares the memory usage and BFS computation time of \sysname and Gemini on \textit{twitter-2010} and \textit{clueweb12}. 
Despite accessing graph data from a remote memory pool, \sysname achieves computation time within $40\%$ of Gemini, which primarily accesses local memory. 
This computation-time gap will further narrow as interconnect bandwidth continues to increase, for example, with 800 Gbps and faster networks~\cite{D-RDMA_CIDR2022,Tassel_SIGCOMM2024}.
In return for this moderate computation overhead, \sysname provides substantial improvements in end-to-end execution time and resource efficiency. 
Unlike Gemini, which reserves the full memory capacity of a fixed set of servers, \sysname allocates memory on demand in proportion to the graph size. 
For smaller graphs such as \textit{twitter-2010}, this reduces memory consumption by an order of magnitude. 
\sysname also provisions compute resources independently of memory capacity, allowing even large graphs such as \textit{clueweb12} to run with only one or two CNs.

\subsection{Factor Analysis}
This section analyzes how individual designs in \sysname contribute to overall performance and validates design choices.

\subsubsection{\textbf{The DM-friendly Graph Store}}
We first study the effect, parameters, and tradeoff in the DM-friendly graph store. 
Experiments are run on one CN to eliminate the multi-CN update effect.

\begin{figure}[tbp]
  \centering
  \begin{minipage}[t]{0.48\linewidth}
    \centering
    \setlength{\abovecaptionskip}{3pt}
    \includegraphics[width=\textwidth]{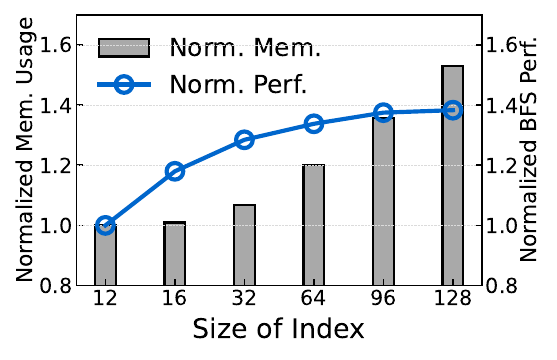}
    \caption{Total MNs' memory usage and BFS performance with different index size (TW, 1 CN)}\label{FIG_EXP_IDX_SIZE}
  \end{minipage}
  \hspace{0em} 
  \begin{minipage}[t]{0.48\linewidth}
    \centering
    \setlength{\abovecaptionskip}{3pt}
    \includegraphics[width=\textwidth]{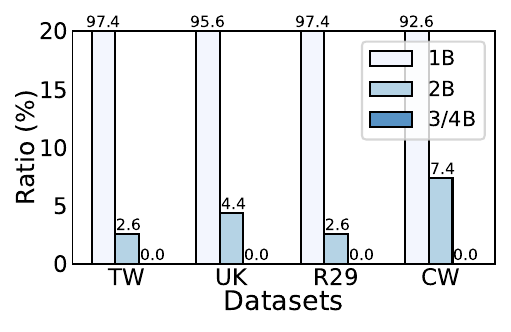}
    \caption{Byte length ratio of segments for out-of-place edge lists when 4 partitions}\label{FIG_EXP_SEG_LEN_BYTE}
  \end{minipage}
\end{figure}

\textbf{Index Structure.} 
As shown in Figure~\ref{FIG_EXP_IDX_SIZE}, 32B and 64B indices deliver substantial speedups with modest memory amplification, while larger indices incur prohibitive memory overhead without proportional benefit. 
Figure~\ref{FIG_EXP_SEG_LEN_BYTE} further shows that 1B suffices to store most segment lengths for various datasets under byte compression, 
suggesting that small indices can capture a substantial amount of segmentation information.

\begin{figure}[tbp]
  \centering
  \setlength{\abovecaptionskip}{3pt}
  \includegraphics[width=0.48\textwidth]{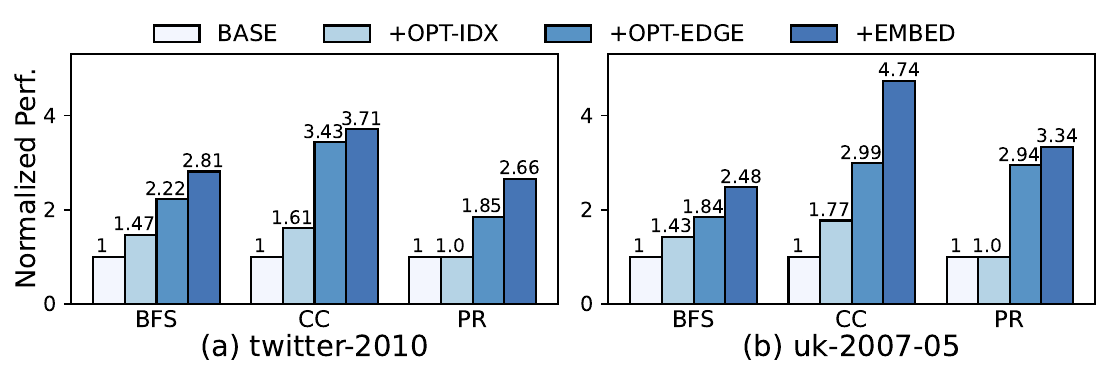}
  \caption{Design effect of retrieval optimizations.}
  \label{FIG_EXP_BREAKDOWN_RETRIEVAL}
\end{figure}

\textbf{Retrieval Efficiency.} 
\sysname's adaptive index scheme and optimizations are designed to improve graph retrieval efficiency.
Figure~\ref{FIG_EXP_BREAKDOWN_RETRIEVAL} shows that applying index/edge-list merging and batching (\texttt{+OPT-IDX, +OPT-EDGE}) and embedding indices (\texttt{+EMBED}) yield 2.48--4.74$\times$ end-to-end speedups, with each optimization contributing substantially.
The one exception is PR: merging and batching of index reads provide little benefit since PR activates all vertices, resulting in naturally sequential access.

\subsubsection{\textbf{Adaptive Update Coordinator}}~\label{sec:exp:breakdown:adap_co_proc}
We evaluate the adaptive update coordinator using 4 CNs to analyze both the impact of re-distribute and the comparison between different methods to perform update propagation.

\begin{figure}[tbp]
  \centering
  \setlength{\abovecaptionskip}{3pt}
    \includegraphics[width=0.48\textwidth]{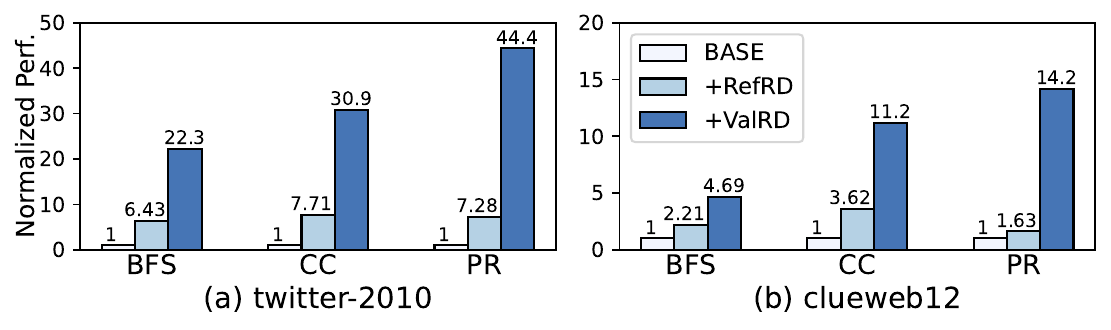}
    \caption{Design effect of two types of RD.}
    \label{FIG_EXP_BREAKDOWN_RD}
\end{figure}

\textbf{Effect of Re-Distribute (RD).} 
Figure~\ref{FIG_EXP_BREAKDOWN_RD} shows the performance impact of two RD schemes. 
\texttt{BASE} performs all updates outside the cached chunk via direct fine-grained remote accesses. 
\texttt{+RefRD} enables \textit{RefRD},
and \texttt{+ValRD} further enables both \textit{RefRD} and \textit{ValRD}.
By converting costly direct remote updates into large sequential RDMA accesses and local updates, both RD schemes substantially improve overall graph processing performance.
\textit{TW} benefits more because it has a larger fraction of cross-partition edges. 
In practice, combining \textit{RefRD} and \textit{ValRD} yields the best performance and using only one pattern is less effective as discussed in §~\ref{sec:design:update:collaborative}.

\begin{figure}[tbp]
  \centering
  \setlength{\abovecaptionskip}{3pt}
    \includegraphics[width=0.48\textwidth]{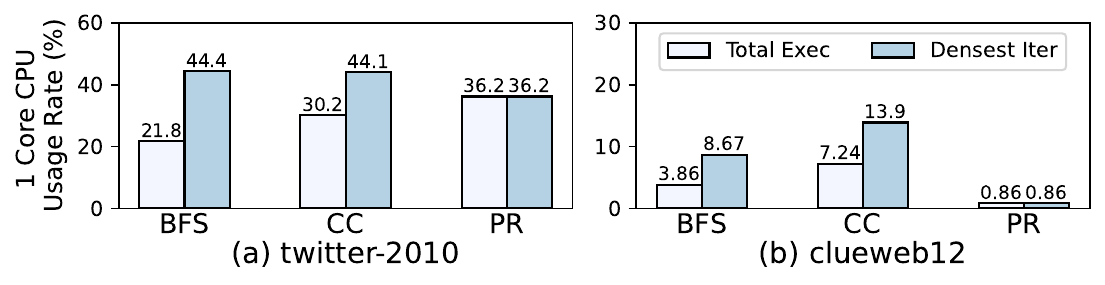}
    \caption{CPU (1 core) usage in MNs for processing ValRD.}
    \label{FIG_EXP_MN_CPU_USAGE}
\end{figure}

\textbf{CPU Usage for \textit{ValRD}.}
Figure~\ref{FIG_EXP_MN_CPU_USAGE} reports CPU usage rate of the single core in MN that processes offloaded \textit{ValRD} items. 
Even for \textit{TW}, which has the highest ratio of edges spanning across chunks, the peak usage rate stays below 44\% in the densest iteration and is even lower throughout the rest of the execution. 
For \textit{CW}, which exhibits better data locality, the usage rate remains below 8\% across the entire execution. 
These results confirm that, after turning fine-grained RDMA operations into large sequential RDMA accesses by RD, processing \textit{ValRD} items (i.e., performing update propagation) places only limited demands on the scarce CPU resources of MNs.

\textbf{Comparison of Update Propagation Methods.}
\sysname includes two methods to perform update propagation.
\textit{Collaborative Update (CO Update)} is designed for dense iterations with limited cache, relying on RD to resolve uncached updates with low overhead. 
\textit{Direct Remote Update (DR Update)} is used in sparse iterations by directly performing updates via remote operations.
To further demonstrate the effectiveness of CO Update, which requires limited cache, we also implement another method for comparison: \textit{Fully-Cached Update (FC Update)}.
In \textit{FC Update}, each CN loads the attributes of all vertices into its local cache without considering cache size limits, performs all updates locally, and merges the locally cached attributes back into remote memory at iteration boundaries.

\begin{figure}[tbp]
  \centering
  \setlength{\abovecaptionskip}{3pt}
    \includegraphics[width=0.48\textwidth]{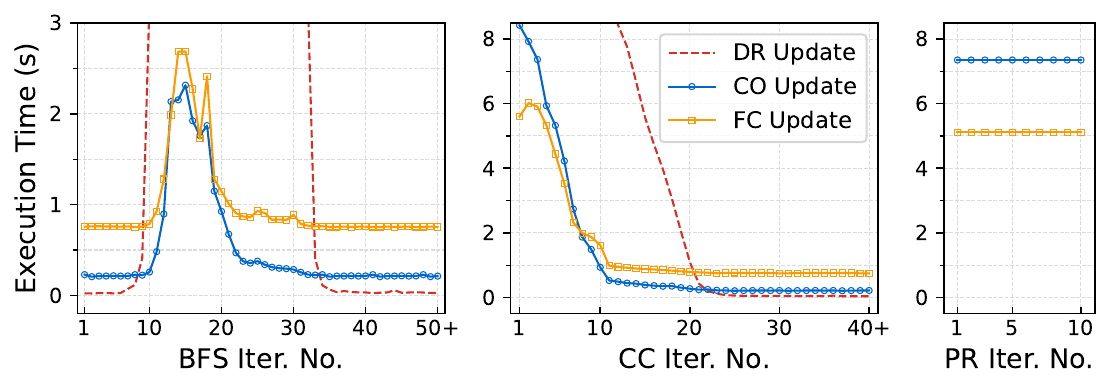}
  \caption{Per-iteration analytics of different methods to perform updates using 4 CNs on \textit{clueweb12}.}
  \label{FIG_EXP_ITERATION_CHOICE}
\end{figure}

Figure~\ref{FIG_EXP_ITERATION_CHOICE} presents a per-iteration analysis of execution time under the three methods.
From the results, we observe that:
(1) The \textit{Collaborative Update} method is efficient. 
Except for a few extremely dense iterations (e.g., early steps of CC or fully-active rounds of PR), CO Update achieves similar or even better performance compared to FC Update while requiring only $\#CN$ times less cache. 
This is because the RD in collaborative update handles uncached updates with low overhead. 
In contrast, FC Update suffers from repeatedly loading and synchronizing full vertex attributes, which incurs substantial bandwidth and synchronization cost.
(2) Adaptive use of \textit{Direct Remote Update} is necessary. 
In many workloads (e.g., BFS or CC, which converge in about 500 iterations on \textit{clueweb12}), a large number of iterations are very sparse, where direct remote updates are more efficient than caching.

\subsubsection{\textbf{Two-Stage Workload Manager}}
\sysname employs a two-stage workload manager: (1) coarse-grained partitioning for fast workload division, and (2) fine-grained runtime re-scheduling to mitigate long-tail effects and improve performance scalability with more threads.

\textbf{Fast Partitioning.} 
Partitioning speed aligns with intuition: tiny-chunk-based partitioning completes in sub-seconds, while partitioning at per-vertex granularity takes tens to hundreds of seconds. 
The one-time preprocessing cost of generating tiny-chunk metadata is comparable to running per-vertex partitioning, meaning our approach is effective whenever partitioning is performed more than once, and not worse even if performed only once. 
In terms of end-to-end execution, which is the ultimate goal of a partitioning strategy, tiny-chunk and per-vertex schemes yield nearly identical BFS computation time on 4 CNs. 
This confirms that \sysname achieves fast partitioning speed without compromising effectiveness.

\begin{figure}[tbp]
  \centering
  \setlength{\abovecaptionskip}{3pt}
  \includegraphics[width=0.48\textwidth]{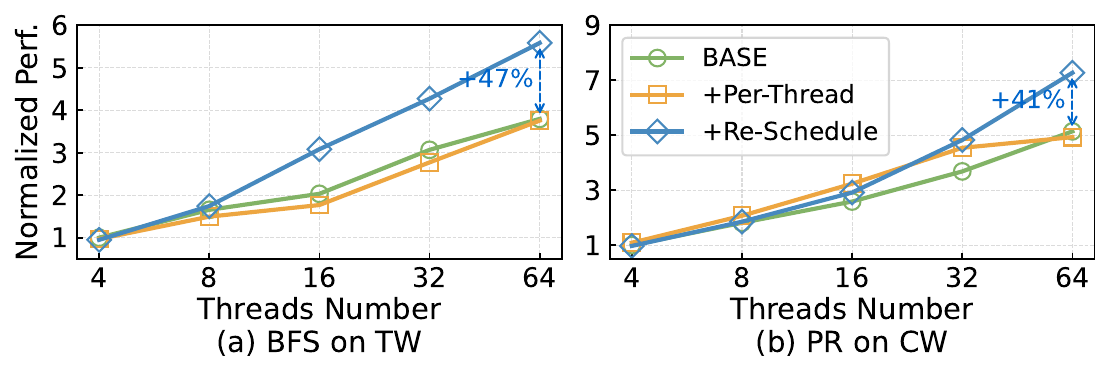}
  \caption{Design effect of \sysname's workload manager. Use 1 CN for \#Threads$\leq$16, 16 threads per CN for \#Threads$\geq$16}
  \label{FIG_EXP_BREAKDOWN_MANAGER}
\end{figure}

\textbf{Scale to More Threads.} 
Figure~\ref{FIG_EXP_BREAKDOWN_MANAGER} reports normalized performance of BFS on TW and PR on CW as thread count increases from 4 to 64. 
\texttt{BASE} uses a centralized per-CN manager without hub-vertex handling. 
\texttt{+Per-Thread} introduces per-thread managers with work stealing, while \texttt{+Re-Schedule} further enables runtime re-scheduling (RS) for hub vertices (full \sysname design).
With RS, performance improves by $47\%$ and $41\%$ at 64 threads, respectively. 
Per-thread management alone does not outperform \texttt{BASE} for BFS on TW, since the small, sparse graph makes per-thread management overhead outweigh its benefits.

\section{Related Work}

\textbf{Large-scale Graph Processing.}
Traditional graph systems handle large graphs in two ways. 
Out-of-core systems~\cite{graphchi_2012,gridgraph_2015,graphene_2017,seraph_fast24,seraph_tos,chunkgraph_2024,oasis_fast25,gem_sigmod,ACGraph_sigmod} store graphs on storage devices to exploit their large capacity, but are limited by I/O bandwidth and single-machine compute power. 
Distributed systems~\cite{gemini_2016,powergraph_2012,powerlyra_eurosys,gluon_2018,Nezha_sigmod} partition graphs across monolithic servers so that each shard fits in memory, but they lack elastic resource management.
FaaSGraph~\cite{FaaSGraph} enables graph processing on serverless platforms. 
Yet, it allocates resources with a fixed CPU/memory ratio, and forces large graphs to span many containers due to the limited capacity in each container. This fragmentation amplifies memory and communication overhead and yields performance inferior to \sysname. 

\textbf{Resource Disaggregation.}
DM has been extensively explored across various layers, including hardware~\cite{lee2021mind,maruf2023tpp,li2023pond,gouk2023memory}, operating systems~\cite{legoos_osdi,infiniswap,fastswap,wang2023canvas}, user-level libraries~\cite{smart_asplos24,wang2022memliner,Patronus_fast23,zhou2022carbink,DM_cache_cohe}, and applications~\cite{motor_osdi,ditto_sosp,HDTX_ATC25,DistVS_NSDI}. 
General resource disaggregation includes storage disaggregation as well~\cite{Cornus_vldb,Volley_eurosys24,storage_disagg_sigmod24}.
\sysname 
is designed for processing graph-structured data on DM and 
can be deployed on diverse infrastructures and leverage low-level optimizations.

\section{Conclusion}
This paper presents \sysname, a graph processing system on DM.
\sysname addresses key challenges in achieving high performance on a practical DM architecture by
a DM-friendly graph store for efficient graph placement and retrieval, 
an adaptive update coordinator for handling complex update propagation,
and a two-stage workload manager for fast and effective load balancing.
Compared to the state-of-the-art graph processing system on DM, \sysname delivers superior system scalability, cache efficiency, and high performance.


\bibliographystyle{ACM-Reference-Format}
\bibliography{DMGraph}










\end{document}